\documentclass[conference,letterpaper]{IEEEtran}

\usepackage{graphics}

\usepackage{amsmath} 
\usepackage{amssymb} 
\usepackage{pifont} 
 \usepackage{cite}
\usepackage{url}
\usepackage{comment}
\usepackage{color, colortbl}
\usepackage{enumerate}

\usepackage[utf8]{inputenc}  
\usepackage[T1]{fontenc}  
\usepackage{curve2e}
\usepackage{fp}

\usepackage{fancyhdr}

\usepackage{graphicx}
\usepackage[export]{adjustbox}

\definecolor{Gray}{gray}{0.9}
\begin{document}
\setlength{\unitlength}{5cm}\linethickness{0.6pt}

\title{Cryptographic Protocol for Multipart Missions Involving Two Independent and Distributed Decision Levels in a Military Context}

\author{\IEEEauthorblockN{Jaouhar Fattahi$^{1,2}$, Mohamed Mejri$^2$, Marwa Ziadia$^2$, Elies Ghayoula$^{3}$, Ouejdene Samoud$^2$ and Emil Pricop$^4$} 
\IEEEauthorblockA{$^1$Defense Research and Development Canada. Valcartier Research Centre. Québec. Canada.}
\IEEEauthorblockA{$^2$Computer Science and Software Engineering Department. Laval University. Québec, Canada.}

\IEEEauthorblockA{$^3$Department of Electrical  and Computer Engineering. 
Laval University. Québec, Canada.}

\IEEEauthorblockA{$^4$Automatic Control, Computers and Electronics Department Petroleum-Gas University of Ploiesti. Romania.}

}



%


\maketitle

\fancypagestyle{plain}{
            \fancyhf{}         
            \fancyfoot[L]{}
            \fancyfoot[C]{}
            \fancyfoot[R]{}
            \renewcommand{\headrulewidth}{0pt}
            \renewcommand{\footrulewidth}{0pt}
}
 
\pagestyle{fancy}{
            \fancyhf{}
            \fancyfoot[R]{}}
\renewcommand{\headrulewidth}{0pt}
\renewcommand{\footrulewidth}{0pt}


\begin{abstract}
In several critical military missions, more than one decision level are involved. These decision levels are often independent and distributed, and sensitive pieces of information making up the military mission must be kept hidden from one level to another even if all of the decision levels cooperate to accomplish the same task. Usually, a mission is negotiated through insecure networks such as the Internet using cryptographic protocols. In such protocols, few security properties have to be ensured. However, designing a secure cryptographic protocol that ensures several properties at once is a very challenging task. In this paper, we propose a new secure protocol for multipart military missions that involve two independent and distributed decision levels having different security levels. We show that it ensures the secrecy, authentication, and non-repudiation properties. In addition, we show that it resists against man-in-the-middle attacks.\\
\end{abstract}


%
\IEEEpeerreviewmaketitle

\section{Introduction}

For many years, military concerns have been focusing on protecting against leakage of sensitive data. A disclosure of sensitive information can be costly in the military field\cite{DBLPLeeKMSGP12, DBLPRamazanaliJVB15,DBLPHurK14,DBLPUcarEOTB16}. The particularity in military operations is that they usually involve several complicit entities, which can potentially form a serious risk. Thus, any military system should be designed based on cryptography to secure data exchange between different military control units. Compulsory usage of cryptographic protocols is the most successful measure to reinforce security especially in military communications that even selective encryption is not enough \cite{wStafrace2010619, w5604602, wKlein2015, wMatsuo2010}. \\

In many military use cases, a military mission requires more than one decision level to be accomplished. It is called a multipart mission. Often, the decision-makers are independent, have different security levels, and communicate with each other remotely throughout an unsecured network such as the Internet. The resulting decision is so an aggregation of distributed decisions. During the same mission, many actors are involved and each one is requested to accomplish only his part without knowing anything about the other parts. Hence, the secrecy inside communications becomes a major challenge as the secrets exchanged between the different actors must be protected. \\

Moreover, considering that a military mission has bad consequences upon failure, every actor must be able to prove, anytime, that he had carried out, or not, a specific action. We talk about the property of non-repudiation here. Besides, since the different actors communicate with each other remotely through an insecure network, every actor must be assured that he communicates with the right actor prior to undertake any action. For that, the authentication property must be maintained in the protocol.\\

It goes without saying that an intruder has an impressive set of capabilities\cite{Dolev1,Ref20,HERZOG200557,Chen1825499} over the net. In fact, he is able to intercept any message, redirect it, or decompose it. Besides, he can encrypt or decrypt any message provided that he possesses the required keys. The only thing he cannot perform is to decrypt an encrypted message when he does not have the decryption key. A secure protocol must closely consider these facts to prevent any man-in-the-middle attack.\\

In this paper, we are proposing a new secure protocol for military missions that require two independent and distributed decision levels having two different security levels to be accomplished. The first level is referred to as the operational level which determines the mission target (i.e. the location to be bombed in the case of a strike operation). This level generally involves both political and operational decision-makers whose decision is issued from a remote command center. The second level is referred to as the the logistic level. This level is concerned with only planning, implementing, and controlling the resources required by the mission. For example, it determines which aircraft is going to be used, the takeoff and landing runway necessary for the mission, and all other related logistic aspects. The logistic level must behave as an active broker between the operational decision-maker and the final mission executor (i.e. the pilot). It receives the operational decision cyphered, appends its logistic decision to it, and delivers the resulting command to the mission executor. This is because any information disclosure related to the operational decision (i.e. the target to be struck) before the mission is finished may very much lead to undesirable outcomes. Hereafter, we demonstrate that our proposed protocol guarantees secrecy, authentication, and non-repudiation. Moreover, it resists against man-in-the-middle attacks. As far as we know, no cryptographic protocol available in the-state-of-the-art guarantees all these security properties at once as our protocol does, in a distributed and independent decision levels' context.

\section{Paper organization}

The paper is organized as follows:

\begin {enumerate}

\item In Section \ref{sec1}, we give the protocol definition (contribution);

\item In Section \ref{sec2}, we give the protocol chronology and we discuss the secrecy and authentication properties (contribution);

\item In Section \ref{sec3}, we show that the protocol resists against man-in-the-middle attacks (contribution);

\item In Section \ref{sec4}, we show that the non-repudiation property is maintained inside the protocol (contribution);

\item In Section \ref{sec5}, we discuss some security issues of our protocol and we compare with other related works, particularly with a chaotic protocol (contribution);

\item in Section \ref{sec6}, we give a short conclusion.

\end {enumerate}

\section{Protocol definition \label{sec1}}

The proposed protocol consists of the following seven steps.
\footnotesize{
$ $\\

$1.~A\rightarrow S:\{A.\{N_A.S.TS_A\}_{k_{a}^{-1}}.S\}_{k_s}$\\

$2.~S\rightarrow B: \{S.B.\{B.N_S.N_A.A\}_{k^{-1}_{s}}\}_{k_b}$\\

$3.~B \rightarrow S: \{B.A.S.N_S\}_{k_s}$\\

$4.~S\rightarrow A: \{\{B.N_A.N_S\}_{k^{-1}_{s}}.A.S\}_{k_a}$\\

$5.~S \rightarrow B: \{\\
\mbox{~~~~~~~~~~~~~~~~~~~~}\{OP\_ACT.S.A.\{A.N_A.S.N_S.B\}_{k^{-1}_s}\}_{k_a}.\{S.N_S\}_{k^{-1}_s}
\\ \mbox{~~~~~~~~~~~~~~~~~~~~} \}_{k_b}$\\

$6.~B \rightarrow A: \{\\ \mbox{~~~~~~~~~~~~~~~~~~~~}\{TS_B.N_S.N_A.LOG\_ACT\}_{k^{-1}_b}.\\ \mbox{~~~~~~~~~~~~~~~~~~~~}\{OP\_ACT.S.A.\{A.N_A.S.N_S.B\}_{k^{-1}_s}\}_{k_a}.B.A.N_A.N_S \\ \mbox{~~~~~~~~~~~~~~~~~~~~}\}_{k_a}$\\

$7.~A\rightarrow B: \{\\
\mbox{~~~~~~~~~~~~~~~~~~~~}\{\hbar(LOG\_ACT.N_A.N_S)\}_{k^{-1}_{a}}.A.B.S.TS_B
\\ \mbox{~~~~~~~~~~~~~~~~~~~~}\}_{k_b}$\\

}

\normalsize
Where: \\

\begin{itemize}
\item $A$: is the mission executor (e.g. a pilot) with a pair of asymmetric keys $(k_a,k^{-1}_a)$.
\item $S$: is the operational decision-maker with a pair of asymmetric keys $(k_s,k^{-1}_s)$.
\item $B$: is the logistic decision-maker with a pair of asymmetric keys $(k_b,k^{-1}_b)$.
\item $N_A$ and $N_S$: are nonces (random numbers).
\item $TS_A$ and $TS_B$: are typed timestamps.
\item $LOG\_ACT$: is the logistic action.
\item $OP\_ACT$: is the operational action.
\item $\hbar$: is a hash function.
\end{itemize}
\normalsize
\section{Protocol chronology and standard security properties\label{sec2}}

In this section we present the protocol chronology. Along with that, we discuss how the protocol preserves the two standard security properties: secrecy and authentication. Let us see this in each step of the protocol.

\begin {enumerate}

\item In the step 1, the mission executor $A$ remotely subscribes on the mission executors' availability list held by $S$. He sends him his identity $A$, as well as  a signed message $\{N_A.S.TS_A\}_{k_{a}^{-1}}$ enclosing a timestamp that ensures $S$ about the freshness of the message, the identity of the receiver $S$, and a nonce $N_A$ (a random number) that will be a  secret shared between both of them and $B$ and used by $S$ and $B$ once $A$ is selected to be assigned to a mission. The signature in $\{N_A.S.TS_A\}_{k_{a}^{-1}}$ authenticates $A$ to $S$. The secrecy of $N_A$ is ensured by the encryption of the whole message with the public key of $S$;\\

\item In the step 2, $S$ urges $B$ that a mission is about to take place involving $A$. Beside his identity $S$ and the identity of $B$, $S$ sends to $B$ a signed message $\{B.N_S.N_A.A\}_{k^{-1}_{s}}$. This signature authenticates $S$ to $B$. Besides, $B$ receives the identity $A$ of the mission executor, the nonce $N_A$ previously issued by $A$, and a nonce $N_S$ issued by $S$. This latter guarantees the freshness of the message. The secrecy of the two nonces is ensured by the encryption of the whole message with the public key of $B$;\\

\item In the step 3, $B$ replies to $S$ by sending him the message $\{B.A.S.N_S\}_{k_s}$. $S$ sees his nonce $N_S$ and gets assured that $B$ is henceforth aware that a mission is going to take place. The authentication of $B$ to $S$ is now established since he is the only to have received $N_S$. The secrecy of $N_S$ is ensured by the encryption of the message with the public key of $S$;\\

\item In the step 4, $S$ urges $A$ that he has been selected to carry out a mission and he is about to be delivered the mission instructions by $B$. These instructions will be discussed in the step 6. The signed message $\{B.N_A.N_S\}_{k^{-1}_{s}}$ authenticates $S$ to $A$. The nonces $N_A$ and $N_S$ ensure the freshness of the upcoming mission. The secrecy of the two nonces is ensured by the encryption of the whole message with the public key of $A$;\\

\item In the step 5, $S$ sends $B$ the message\\ 
$\{\\
 \mbox{~} \{OP\_ACT.S.A.\{A.N_A.S.N_S.B\}_{k^{-1}_s}\}_{k_a}.\{S.N_S
 \}_{k^{-1}_s}\\ \}_{k_b}$. \\

The enclosed message $\{OP\_ACT.S.A.\{A.N_A.S.N_S.B\}_{k^{-1}_s}\}_{k_a}$ is not intelligible for $B$ (encrypted with the public key of $A$) because it contains the mission details related to the operational decision-maker (i.e. $OP\_ACT$) that must be shared between $S$ and $A$ only. The encryption with the public key of $A$ guarantees the secrecy of theses details. The enclosed signed message $\{S.N_S\}_{k^{-1}_s}$ ensures $B$ that the whole messages truly comes from $S$. The nonce $N_S$ ensures the freshness of the message;\\

\item In the step 6, $B$ sends a three-part encrypted message to $A$. The first part is the signed message $\{TS_B.N_S.N_A.LOG\_ACT\}_{k^{-1}_b}$ that contains the details  concerning the logistic part of the mission (i.e. $LOG\_ACT$). It contains also the timestamp that guarantees the freshness of the message as well as the nonces $N_A$ and $N_S$ that ensure that this step is connected to the previous ones. The signature ensures $A$ that the received message truly comes from $B$. The second part $\{OP\_ACT.S.A.\{A.N_A.S.N_S.B\}_{k^{-1}_s}\}_{k_a}$ is exactly what $B$ received from $S$ during the previous step. It contains the operational mission details (i.e. $OP\_ACT$) as well as a signed message $\{A.N_A.S.N_S.B\}_{k^{-1}_s}$ that guarantees that it truly comes from $S$. The last part $B.A.N_A.N_S$ contains the nonces $N_S$ and $N_B$ as well as the identity of the sender and the receiver. The duplication of the nonces in the three parts guarantees that they have the same freshness and are related to the same mission. Now $A$ has the complete and authentic information to execute the mission;\\

\item In the step 7, $A$ acknowledges $B$ that he receives all the required instructions to carry out the mission. $B$, upon receiving the signed message $\{\hbar(LOG\_ACT.N_A.N_S)\}_{k_{a}^{-1}}$, gets assured that the message truly comes from $A$ and the freshness of this message is guaranteed by the nonces. The timestamp $TS_B$ guarantees the freshness of the whole the message. \\

\end {enumerate}

\section{Resistance against man-in-the-middle attacks\label{sec3}}

A man-in-the-middle attack\cite{Lowe1995SMC2,DBLPCashKT16,Anderson2009,Anada2010,klpo7846989,MITM4768661} arises when an intruder succeeds to send an illegitimate message to a regular agent and this latter naively accepts it because there is no way for him to know whether it is legitimate or not. The illegitimate message issued by the intruder is a message intercepted during previous runs of the protocol or within previous steps of the running instance of it. When we designed our protocol, we bore in mind this famous sort of attack. Therefore, we designed it in such a way that any two messages exchanged in the protocol do not overlap by carefully choosing the position\cite{AbadiTyping,DBLjournaloDebbabiDMM03} of agent identities, signed messages, nonces, etc. in every single message. Besides, we made sure that the freshness of all the messages, when received, could easily be verified by the receiver. Let us see this in each step of the protocol.\\

\begin {enumerate}

\item In the step 1: $S$ receives the message $\{A.\{N_A.S.TS_A\}_{k_{a}^{-1}}.S\}_{k_s}$. Upon decryption, $S$ observes that the identity of the sender $A$ appears first, then appears the signed message $\{N_A.S.TS_A\}_{k_{a}^{-1}}$, then his identity $S$. This structural pattern does not overlap with any of the other messages exchanged in the protocol. So, he concludes that the message is a regular one and was truly originated by $A$. Besides, the presence of the timestamp and the nonce guarantee that the message is fresh and not an old one intercepted and resent by the intruder;\\

\item In the step  2: $B$ receives the message $\{S.B.\{B.N_S.N_A.A\}_{k^{-1}_{s}}\}_{k_b}$. Upon decryption, $B$ observes that the identity of the sender $S$ appears first, then appears his identity $B$, then the signed message $\{B.N_S.N_A.A\}_{k^{-1}_{s}}$. This structural pattern does not overlap with any of the other messages exchanged in the protocol. So, he concludes that the message is a regular one and was truly originated by $S$. Besides, he can verify that the pair $(N_S,N_A)$ has never been used before and hence make sure that the message is fresh and not an old one intercepted and resent by the intruder;\\

\item In the step  3: $S$ receives the message $\{B.A.S.N_S\}_{k_s}$. Upon decryption, $S$ observes that the identity of the sender $B$ appears first, then appears the identity of the mission executor $A$, then appears his identity $S$, then appears a nonce $N_S$. This structural pattern does not overlap with any of the other messages exchanged in the protocol. So, he concludes that the message is a regular one and was truly originated by $B$. Besides, he can verify that the message is fresh by verifying that the nonce $N_S$ is the same one he generated in the previous step;\\

\item In the step  4: $A$ receives the message $\{\{B.N_A.N_S\}_{k^{-1}_{s}}.A.S\}_{k_a}$. Upon decryption, he observes that the signed message $\{B.N_A.N_S\}_{k^{-1}_{s}}$ appears first, then his identity $A$, then the identity of the operational decision-maker $S$. This structural pattern does not overlap with any of the other messages exchanged in the protocol. So, he concludes that the message is a regular one and was truly originated by $S$. Besides, he can verify that the pair of nonces $(N_S,N_A)$ has never been used before and hence he makes sure that the message is fresh and not an old one intercepted and resent by the intruder;\\

\item In the step  5: $B$ receives the message \\
$\{\\
\{OP\_ACT.S.A.\{A.N_A.S.N_S.B\}_{k^{-1}_s}\}_{k_a}.\{S.N_S\}_{k^{-1}_s}
\\ \}_{k_b}$.\\ 

Upon decryption, $B$ observes that the signed message $\{S.N_S\}_{k^{-1}_s}$ including the identity of the operational decision-maker $S$ and its nonce $N_S$ appears in the last position. This structural pattern does not overlap with any of the other messages exchanged in the protocol. So, he concludes that the message is a regular one and was truly originated by $S$. Besides, he can verify that the nonce $N_S$ is the same one received from $S$ in the step 3 to make sure that the message is fresh and not an old one intercepted and resent by the intruder;\\

\item In the step  6: $A$ receives the message\\ 

$\{\\ \mbox{~}\{TS_B.N_S.N_A.LOG\_ACT\}_{k^{-1}_b}.\\ \mbox{~}\{OP\_ACT.S.A.\{A.N_A.S.N_S.B\}_{k^{-1}_s}\}_{k_a}.\\ \mbox{~}B.A.N_A.N_S \\ \mbox{}\}_{k_a}$.\\ 

Upon decryption, $A$ observes that the messages ends with the string $B.A.N_A.N_S$, in which appear the identity of the logistic decision-maker $B$, then his own identity $A$, then his own nonce $N_A$, then the nonce $N_S$ of the operational decision maker $S$. This structural pattern does not overlap with any of the other messages exchanged in the protocol. So, he concludes that the message is a regular one and was truly originated by $B$. Besides, he can verify that the nonce $N_A$ is the same one he generated in Step 1 and hence he gets assured that the message is fresh and not an old one intercepted and resent by the intruder;\\

\item In the step  7: $B$ receives the message \\
$\{\\
 \mbox{~}\{\hbar(LOG\_ACT.N_A.N_S)\}_{k_{a}^{-1}}A.B.S.TS_B
\\ \}_{k_b}$. \\

Upon decryption, $B$ observes that the message ends with a timestamp. This structural pattern does not overlap with any of the other messages exchanged in the protocol. So, he concludes that the message is a regular one and was truly originated by the mission executor $A$. The timestamp, as well as pair of nonces $(N_S,N_A)$, ensure that the message is fresh and not an old one intercepted and resent by the intruder.

\end {enumerate}

Hence, we conclude that the protocol resists against the man-in-the-middle attacks.

\section{About the non-repudiation property\label{sec4}}

The repudiation occurs when an agent can deny an action he carried out in the protocol without being able to prove the opposite from the protocol rules\cite{DBLPLiuV10,SchickR12, WuZXX13}. This cannot happen in the proposed protocol. In fact, in the previous section, we have shown that every message exchanged in the protocol has a unique structural pattern that characterizes it and prevents it from overlapping with any other message. Besides, we have shown that the freshness is always maintained. Moreover, in each step of the protocol, the identity of the sender appears in the sent message, either explicitly, or in an enclosed signed message, or in both. For example, in the step 3, the identity of the sender, who is the logistic decision-maker $B$, explicitly appears in the sent message $\{B.A.S.N_S\}_{k_s}$. In the same vein, in the step 4, the identity of the sender, who is the operational decision-maker, explicitly appears in the sent message $\{\{B.N_A.N_S\}_{k^{-1}_{s}}.A.S\}_{k_a}$ as well as in the enclosed signed message $\{B.N_A.N_S\}_{k^{-1}_{s}}$. In addition, all the encryptions use asymmetric keys. The non-repudiation property is so guaranteed in the protocol.

\section{Discussion and comparison with related works\label{sec5}}

In this paper, we have proposed a new protocol for multipart military missions involving two independent and distributed decision levels. A decision-maker could be either a physical military staff or an intelligent software application. We have shown that it verifies the following security properties: secrecy, authentication and non-repudiation. We have also shown how and why this protocol resists against any man-in-the-middle attack. Our protocol is designed to be a tagged protocol. In fact, tagging \cite{marwaBlanchetPodelskiTCS04} is a syntactic annotation added to protocol messages to fill the verification procedure and the protocol gaps. It is a simple technique that facilitates the decoding of incoming messages, giving a good protocol design. Tagged protocols are indeed less prone to errors. This is mean that under attacks, a tagged version could be more secure than an untagged one. Tagging prevents also from type-flaw attacks\cite{marwaHeather00} by preventing the substitution problem. The substitution problem is a hard problem in cryptographic protocols and is the source of well-known man-in-the-middle attacks. It consists in substituting a regular message by another message intercepted by an intruder in previous executions of the protocol and taking advantage of the ignorance of the receiver of some components of that message and the fact that it could not perform any verification upon. For instance, let us have a look at a flawed version of the Needham-Schroeder public-key protocol\cite{Lowe1995SMC2,Lowe96breakingandSMC2,Schneider96usingcspSMC2} given by Table \ref{NSLVar}. In this protocol the nonces $N_a$ and $N_b$ are supposed to be two secrets shared between the regular agents $A$ and $B$ only. To attack this protocol, an intruder $I$ executes the first step, but instead of sending a regular nonce to the agent $B$ he sends his identity. The agent $B$ when receives $\{A.\textit{I}\}_{k_b}$ wrongly believes that the identity \textit{I} is a nonce sent by the agent $A$. Then, he executes the second step of the protocol by sending the message $\{\textit{I}.N_b.B\}_{k_a}$ to $A$. Upon reception, the agent $A$ wrongly believes that he is receiving a message from an agent $I$ executing the first step of the protocol where the concatenated string $N_b.B$ is his supposed nonce $N_i$. So, he executes the second step of the protocol and sends the message $\{N_b.B.N_A.A\}_{k_i}$ to $I$. Ultimately, the intruder $I$ has just to decrypt that message and ends up revealing the secrets $N_b$ and $N_b$. This attack is given by Fig. \ref{fig1}.

\begin{table}[h]
\caption{A Flawed Version of the Needham-Schroeder Public-Key Protocol}\label{NSLVar}
\begin{center}
\scalebox{1.0}{
\begin{tabular}{lllll}
  $1.$ & $A$ & $\longrightarrow$ & $B:$ & $\{A.N_a\}_{k_b} $ \\
  $2.$ & $B$ & $\longrightarrow$ & $A:$ & $\{N_a.N_b.B\}_{k_a}$ \\
  $3.$ & $A$ & $\longrightarrow$ & $B:$ & $\{N_b\}_{k_b}$
\end{tabular}
}
\end{center}

\end{table}


\footnotesize{
\setlength{\unitlength}{1mm}
\thicklines
\begin{figure}[h]

   \centering
\caption[]{\label{FailleNSDescAnnexeLong} A MAN-IN-THE-MIDDLE ATTACK ON THE NEEDHAM-SCHROEDER PROTOCOL DUE TO THE SUBSTITUTION PROBLEM}\label{fig1}
\scalebox{.95}{
\begin{picture}(40,20)(25,0)
\put(0,-1){\textit{\textcolor{red}{I}}}
\put(10,8){$\{A.\underbrace{\textit{\textcolor{red}{I}}}_{\textcolor{red}{N_{i(A)}}}\}_{k_b}$}
\put(1,0){
\linethickness{0.25mm}   \vector(1,0){40}
}
\put(43,-1){\textit{B}}
\put(43,-3){
{\linethickness{0.25mm}\vector(0,-1){15}}
}
\put(43,-23){\textit{B}}
\put(47,-22){\linethickness{0.25mm}\vector(1,0){40}}
\put(67,-15){$\{\textcolor{red}{I}.\underbrace{N_b.B}_{\textcolor{red}{N_{i}}}\}_{k_a}$}
\put(89,-23){\textit{A}}
\put(89,-40){\textit{A}}
\put(86,-39){
\linethickness{0.25mm}\vector(-1,0){84}
}
\put(10,-35){$\{N_b.B.N_a.A\}_{\textcolor{red}{k_{i}}}$}

\put(0,-40){\textit{\textcolor{red}{I}}}

\put(10,-45){\textit{I}: is the intruder}
\end{picture}
}
\\\hspace{\linewidth}
\\\hspace{\linewidth}
\\\hspace{\linewidth}
\\\hspace{\linewidth}
\\\hspace{\linewidth}
\\\hspace{\linewidth}
\\\hspace{\linewidth}
\\\hspace{\linewidth}
\\\hspace{\linewidth}
\\\hspace{\linewidth}
\end{figure}
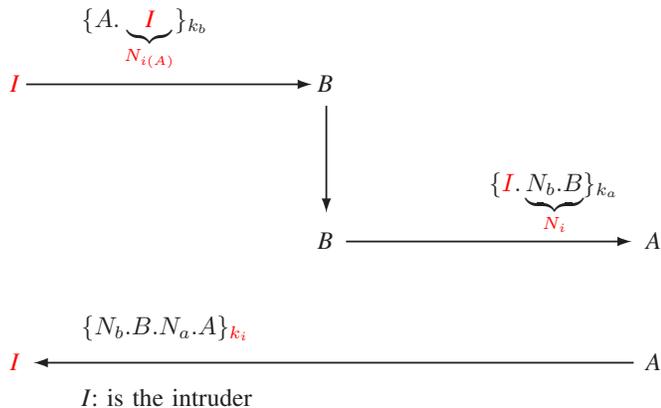
}
\normalsize
$ $\\

This attack could have been prevented if, in the first step the sent message had been $\{N_a.B.A\}_{k_b}$, and in the second step the sent message had been $\{N_a.N_b.B.A\}_{k_a}$. Should this have been the case, the identity of the receiver would have appeared in the second position in the first sent message, last in the second sent message, and wouldn't have appeared at all in the third message. This would have been enough to make every encrypted pattern unique in the protocol and would have precluded any message from overlapping with another one, and hence, the substitution problem could have been avoided. This solution that we are proposing to correct this protocol is given in Table \ref{NSLVarTagged}. 

\begin{table}[h]
\caption{A Tagged and Corrected Version of the Needham-Schroeder Public-Key Protocol}\label{NSLVarTagged}
\begin{center}
\scalebox{1.0}{
\begin{tabular}{lllll}
  $1.$ & $A$ & $\longrightarrow$ & $B:$ & $\{N_a.B.A\}_{k_b} $ \\
  $2.$ & $B$ & $\longrightarrow$ & $A:$ & $\{N_a.N_b.B.A\}_{k_a}$ \\
  $3.$ & $A$ & $\longrightarrow$ & $B:$ & $\{N_b\}_{k_b}$
\end{tabular}
}
\end{center}

\end{table}

The tag information is in fact added by honest agents to precise the intended type of the message and the receiver uses it to identify the message. Hence, tags ensure that a message which has originally a given type cannot not be interpreted as having another type, which is the type-flaw attack definition. Moreover, tagging schemes are used for a decidability proof \cite{marwaRamanujam2003}. Therefore, tagging prevents the unification of different encrypted subterms and yields decidability. Tagging enforces the proof termination \cite{marwaBlanchetPodelskiTCS04}, as well.\\

Back to our protocol, we have made sure that every sent message is tagged so that it is always differentiable from all other messages, and hence, the receiver can always perform the necessary verifications to make certain, and in a sensible way, that the message he is receiving is a legitimate one and has not been intercepted and diverted by an intruder, as explained above in each step of the protocol definition. The substitution problem could never happen in our protocol.\\

In an upcoming work, we are going to give the formal proof of the security properties assured by the proposed protocol using our recent theory called the theory of witness-functions\cite{ChapterJFSpringer}. This theory proposes a series of functions that have been successfully experimented on lots protocols and were powerful either in detecting security flaws or in proving security properties inside cryptographic protocols\cite{trustcomFattahiMP16,DBLFattahiMGP16,TheseJF, WitnessArt1, SMCJaouharECAI2}. Originally, they were designed to prove secrecy inside protocols by proving that the level of security of every atomic message exchanged in the protocol does not decrease during its life cycle. Then, they were extended to cover the property of authentication\cite{trustcomFattahiMP16,DBLFattahiMGP16} by setting few sufficient conditions such that, if met, the protocol is declared correct for authentication. On the other hand, we will closely examine the behavior of the protocol inside a multi-protocol environment\cite{rtyArapinis2015,DBLPCortierD09} where several protocols run simultaneously. In fact, a given protocol can be proven secure when it runs alone, but it might be proven insecure when it runs  simultaneously with other protocols using the same encryption keys. Another issue that should be looked after is the algebraic properties that the used cryptographic protocols primitives could have. Such properties could engender some annoying flaws. An example of algebraic properties may be the homomorphic property\cite{DBLPFioreMNP16,DBLPDahlD14} where $\{m_1.m_2\}_{k}$ is equivalent to  $\{m_1\}_{k}.\{m_2\}_{k}$ under the equational theory\cite{SMCJaouharECAI2,DreierDKS17,ChadhaCCK16}.

\section{Conclusion\label{sec6}}

In this paper, a new protocol for multipart military missions involving two independent and distributed decision levels, operational and logistic, has been presented. We have exhibited the protocol chronology and explained how it fulfills the secrecy and authentication properties in each step of its definition. We have discussed as well how it guarantees the non-repudiation property. We have also examined the resistance of this protocol against the man-in-the-middle attack and discussed the related security concerns. Finally, we introduced to an upcoming work which consists in giving the formal proof for the protocol security correctness using the theory of witness-functions. In the same vein. we intend to extend our protocol so that it can cope with more than two independent and distributed decision levels. We will also keep an eye on using variations of this protocol in cloud computing where multipart tasks are intensively used and security is highly required.\\

\bibliographystyle{ieeetr}
\bibliography{Ma_these} 
\section*{Notice}
\small{
© 2017 IEEE. Personal use of this material is permitted. Permission from IEEE must be obtained for all other uses, in any current or future media, including reprinting/republishing this material for advertising or promotional purposes, creating new collective works, for resale or redistribution to servers or lists, or reuse of any copyrighted component of this work in other works.
}

\end{document}